\documentclass[conference]{IEEEtran}
\usepackage{amsmath,amsfonts}
\usepackage{graphicx,amssymb}
\usepackage{algorithmic,algorithm}
\usepackage{amsthm,dsfont}
\usepackage{subfigure,url}
\usepackage[noadjust]{cite}
\usepackage{subfigure}

\theoremstyle{plain}

\newcounter{longequ}[longequ]
\addtocounter{longequ}{8}

\allowdisplaybreaks

\newcommand{\Tr}{{\mathrm{tr}}}
\newcommand{\Nt}{{N_\mathrm{t}}}
\IEEEoverridecommandlockouts
\begin{document}
%
\title{Optimal Beamforming for MISO Communications via Intelligent Reflecting Surfaces}
\author{\IEEEauthorblockN{Xianghao Yu, Dongfang Xu, and Robert Schober}
\IEEEauthorblockA{Friedrich-Alexander-Universit\"{a}t Erlangen-N\"{u}rnberg, Germany\\
Email: \{xianghao.yu, dongfang.xu, robert.schober\}@fau.de}
\thanks{The work of X. Yu was supported by the Alexander von Humboldt Foundation.}
}


%

\IEEEspecialpapernotice{(Invited Paper)}

\maketitle

\begin{abstract}
Wireless communications via intelligent reflecting surfaces (IRSs) has received considerable attention from both academia and industry. In particular, IRSs are able to create favorable wireless propagation environments with typically low-cost passive devices.
While various IRS-aided wireless communication systems have been investigated in the literature, thus far, the optimal design of such systems is not well understood.
In this paper,  IRS-assisted single-user multiple-input single-output (MISO) communication  is investigated. To maximize the spectral efficiency, a branch-and-bound (BnB) algorithm is proposed to obtain globally optimal solutions for both the active and passive beamformers at the access point (AP) and the IRS, respectively. 
Simulation results confirm the effectiveness of deploying IRSs in wireless systems. Furthermore, by taking the proposed optimal BnB algorithm as the performance benchmark, the optimality of  existing design algorithms is investigated.
\end{abstract}

\IEEEpeerreviewmaketitle

\section{Introduction}
Various technologies have been leveraged for realizing  enhanced mobile broadband (eMBB) in the upcoming 5G wireless communication networks, e.g., deploying large-scale antenna arrays, network densification with small cells, and uplifting the carrier frequency to extremely high frequencies (EHF) \cite{6824752}.
However, additional cost and power consumption are inevitably incurred by deploying more antenna elements, access points (APs), and radio frequency (RF) chains at EHF. Therefore, new cost-effective paradigms that are both spectral- and energy-efficient are needed for future wireless communication systems \cite{zhang2019multiple}.

Because of their ability to control the propagation directions of electromagnetic (EM) waves, intelligent reflecting surfaces (IRSs) have been recently introduced in wireless communication systems  \cite{di2019smart}.
One of the advantages of deploying IRSs in wireless systems is that they are typically composed of low-cost \emph{passive} devices, e.g., phase shifters and dipoles \cite{8466374}. Moreover, the artificial thin films of IRSs can be readily implemented on the facades of infrastructures. Therefore, IRSs are promising enablers for economical and energy-efficient future wireless communication systems \cite{8910627}.
Nevertheless, to fully exploit the potential of IRSs, they have to be properly designed and integrated with conventional communication techniques, such as power allocation and beamforming.

There are several recent works on the design of IRS-assisted wireless communication systems \cite{8811733,tang2019joint,8855810,xu2019resource,zhang2019capacity,yu2019enabling,zhou2019intelligent,yu2019robust}.
A major obstacle in optimizing IRS-assisted wireless systems are the highly non-convex unit modulus constraints (UMCs) associated with the phase shifter implementation. 
The UMCs were tackled via semidefinite relaxation (SDR) 
\cite{8811733,tang2019joint}, whose 
performance was then further improved  via manifold optimization \cite{8855810,xu2019resource}.
Element-wise block coordinate descent (BCD) was employed to handle the UMCs by optimizing one phase shifter at a time
\cite{zhang2019capacity,yu2019enabling}. 
In addition, majorization minorization (MM) and successive convex approximation (SCA) techniques were adopted to deal with the UMCs in \cite{yu2019enabling,zhou2019intelligent} and \cite{yu2019robust}, respectively. 
However, none of the existing optimization algorithms is guaranteed to yield an optimal solution for the unit modulus constrained problems typical for IRS-assisted wireless systems. More importantly, it is difficult to verify the degree of optimality of existing suboptimal algorithms without the globally optimal solution.

In this paper, we consider point-to-point multiple-input single-output (MISO) communication via an IRS implemented by programmable phase shifters. 
To maximize the spectral efficiency, both the active beamformer at the AP and the passive beamformer at the IRS are jointly optimized.
A branch-and-bound (BnB) algorithm is proposed to solve the unit modulus constrained problem.
Unlike the existing results in \cite{8811733,tang2019joint,8855810,xu2019resource,zhang2019capacity,yu2019enabling,zhou2019intelligent,yu2019robust}, the proposed BnB algorithm guarantees the globally optimal solutions for the beamformers at the AP and the IRS. 
Promisingly, our simulation results confirm that the deployment of IRSs significantly improves the spectral efficiency of the considered system. 
More importantly, by taking the proposed BnB algorithm as the performance benchmark, the existing low-complexity manifold optimization-based algorithm in \cite{8855810} is shown to be near-optimal.

\emph{Notations:} The imaginary unit of a complex number is denoted by $\jmath=\sqrt{-1}$. Matrices and vectors are denoted by boldface capital and lower-case letters, respectively. $\mathbb{C}^{m\times n}$ denotes the set of all $m\times n$ complex-valued
matrices. 
$\mathbf{1}_m$ is the $m$-dimensional all-one vector. 
The $i$-th element of vector $\mathbf{a}$ is denoted by $a_i$.
$\mathbf{A}^*$, 
$\mathbf{A}^T$, and $\mathbf{A}^H$ stand for the conjugate, transpose, and conjugate transpose of matrix $\mathbf{A}$, respectively. 
The $\ell_2$-norm of vector $\mathbf{a}$ is expressed as $\left\Vert\mathbf{a}\right\Vert_2$.
$\mathrm{diag}(a_1,\dots , a_n)$ denotes
a diagonal matrix whose diagonal entries are $a_1,\dots, a_n$, while 
$\mathrm{Diag}(\mathbf{A})$ represents a vector whose elements are extracted from the diagonal elements of matrix $\mathbf{A}$.
$\mathbf{A}\succeq0$ indicates that $\mathbf{A}$ is a positive semidefinite (PSD) matrix. $\mathbb{E}[\cdot]$ represents statistical expectation.
The real and imaginary parts of a complex number are denoted by $\Re(\cdot)$ and $\Im(\cdot)$, respectively. 
The operation $\mathrm{unt}(\mathbf{a})$ forms a vector whose elements are $\frac{a_1}{|a_1|},\dots,\frac{a_n}{|a_n|}$.
The Hadamard product between two matrices is denoted by $\circ$. 

\section{System Model and Problem Formulation}
In this section, the signal model of the considered IRS-assisted single-user MISO communication system is first presented. Then, the spectral efficiency maximization problem is formulated, followed by a discussion of existing algorithms.
\subsection{Signal Model}
\begin{figure}
	\centering\includegraphics[width=6cm]{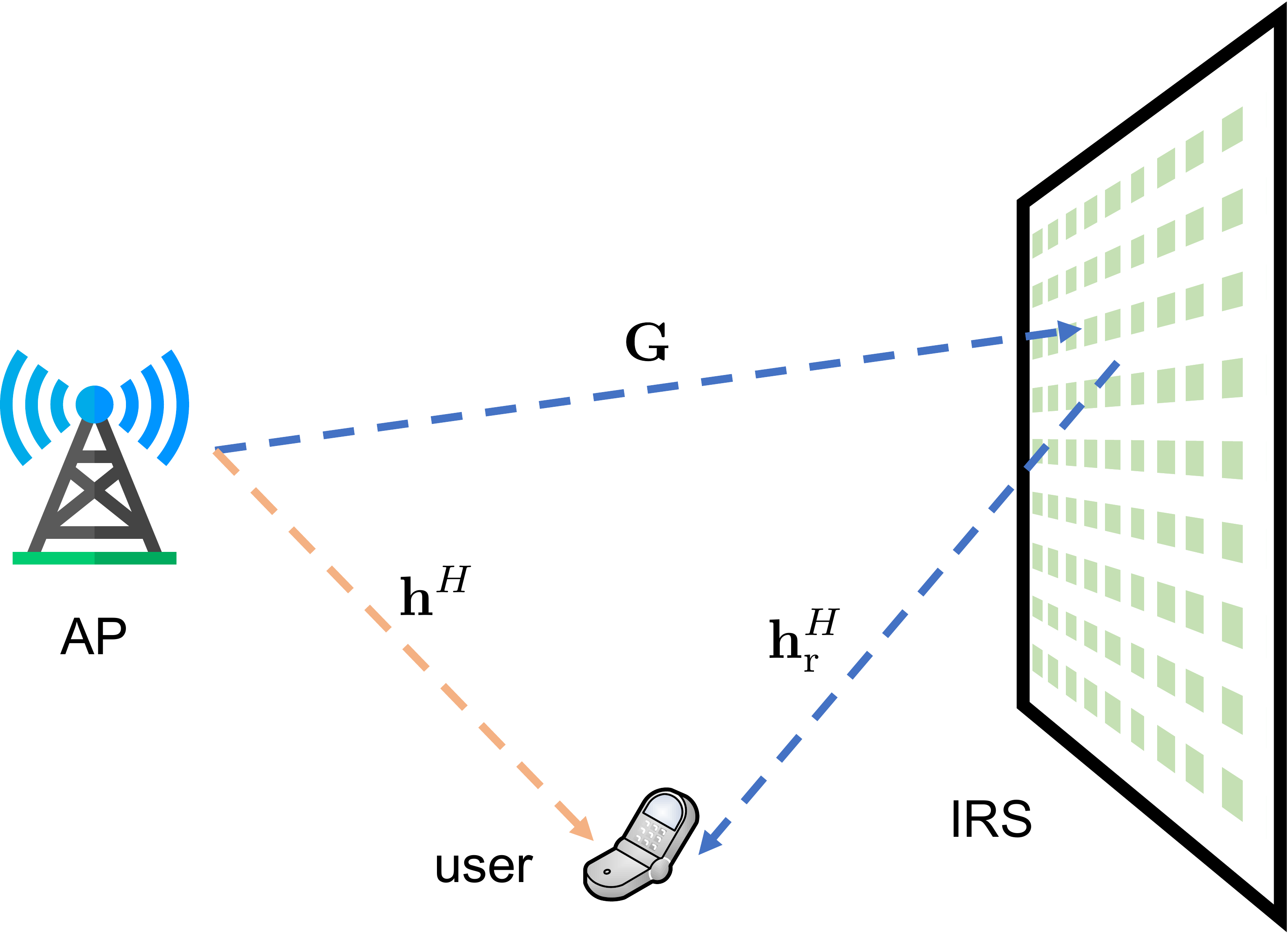}
	\caption{An IRS-assisted point-to-point MISO wireless communication system.}\label{model}
\end{figure}
Consider a point-to-point MISO communication system, which consists of an $\Nt$-antenna AP, a single-antenna user, and an IRS implemented by $M$ configurable phase shifters, as shown in Fig. \ref{model}.
We assume a quasi-static flat-fading channel model and perfect channel state information (CSI) knowledge at both the AP and the IRS\footnote{In practice, CSI can be accurately and efficiently obtained via various channel estimation techniques \cite{wang2019channel}. The results in this paper serve as theoretical performance upper bounds for the considered system, and  provide guidelines for the system design  when the CSI is not perfectly known.}.
Hence, the received signal at the user is given by
\begin{equation}
y = \left(\mathbf{h}_\mathrm{r}^H\mathbf{\Phi}\mathbf{G}+\mathbf{h}^H\right)\mathbf{f}x+n,
\end{equation}
where $\mathbf{h}_\mathrm{r}\in\mathbb{C}^{M\times1}$ is the  channel vector from the IRS to the user, $\mathbf{h}\in\mathbb{C}^{\Nt\times1}$ represents the direct link from the AP to the user, and the channel matrix from the AP to the IRS is denoted as $\mathbf{G}\in\mathbb{C}^{M\times\Nt}$.
The active beamforming vector at the AP and the passive beamforming matrix at the IRS are denoted by  $\mathbf{f}\in\mathbb{C}^{\Nt\times1}$ and $\mathbf{\Phi}=\mathrm{diag}(e^{\jmath \theta_1},e^{\jmath \theta_2},\dots,e^{\jmath \theta_M})$, respectively, where $\theta_i$ denotes the phase shift of the $i$-th  element of the IRS.
%
The transmitted signal is denoted by $x$, where $\mathbb{E}[|x|^2]=1$ without loss of generality, and
$n$ is additive complex Gaussian noise with variance $\sigma^2$.

\subsection{Problem Formulation}
In this paper, our goal is to maximize the achievable spectral efficiency by optimizing both the active beamforming vector $\mathbf{f}$ and passive beamforming matrix $\mathbf{\Phi}$. The spectral efficiency is given by
\begin{equation}
R=\log_2\left(1+\frac{\left|(\mathbf{h}_\mathrm{r}^H\mathbf{\Phi}\mathbf{G}+\mathbf{h}^H)\mathbf{f}\right|^2}{\sigma^2}\right),
\end{equation}
and the resulting optimization problem is formulated as
\begin{equation}\label{eq3}
\begin{aligned}
&\underset{\mathbf{f},\mathbf{\Phi}}{\mathrm{maximize}} && \left|\left(\mathbf{h}_\mathrm{r}^H\mathbf{\Phi}\mathbf{G}+\mathbf{h}^H\right)\mathbf{f}\right|^2\\
&\mathrm{subject\thinspace to}&&\left\Vert\mathbf{f}\right\Vert^2_2\le P\\
&&&\mathbf{\Phi}=\mathrm{diag}\left(e^{\jmath \theta_1},e^{\jmath \theta_2},\dots,e^{\jmath \theta_M}\right),
\end{aligned}
\end{equation}
where $P>0$ is the given maximum transmit power.



Similar to the derivation steps in \cite[Eqs. (15)-(18)]{8811733}, the optimization problem in \eqref{eq3} can be reformulated as
\begin{equation}\label{eq4}\mathcal{P}_1:\quad
\begin{aligned}
&\underset{\mathbf{v}\in\mathbb{C}^{M+1}}{\mathrm{minimize}} && f(\mathbf{v})=\mathbf{v}^H\mathbf{Rv}\\
&\mathrm{subject\thinspace to}&&|v_i|=1,\quad i=1,2,\dots,M+1,
\end{aligned}
\end{equation}
where $\mathbf{v}=[\mathbf{x}^T,t]^T$, $\mathbf{x}=\left[e^{\jmath \theta_1},e^{\jmath \theta_2},\dots,e^{\jmath \theta_M}\right]^H$, $t\in\mathbb{C}$, and
\begin{equation}\label{eq5}
\mathbf{R}=-\begin{bmatrix}
\mathrm{diag}\left(\mathbf{h}_\mathrm{r}^H\right)\mathbf{GG}^H\mathrm{diag}\left(\mathbf{h}_\mathrm{r}\right)&
\mathrm{diag}\left(\mathbf{h}_\mathrm{r}^H\right)\mathbf{G}\mathbf{h}\\
\mathbf{h}^H\mathbf{G}^H\mathrm{diag}\left(\mathbf{h}_\mathrm{r}\right)&0
\end{bmatrix}.
\end{equation}
Note that optimization variable $\mathbf{v}$ in $\mathcal{P}_1$ is composed of an auxiliary variable $t$ and the phase shifts $\{\theta_i\}_{i=1}^M$. Once the optimal solution for $\mathbf{v}$ in $\mathcal{P}_1$ is obtained, the optimal passive beamforming matrix $\mathbf{\Phi}$ can be recovered from the first $M$ elements of $\mathbf{v}$, and therefore the corresponding active beamforming vector $\mathbf{f}$ at the AP is optimally given by the maximum ratio transmission (MRT) strategy, i.e., 
\begin{equation}\label{eq6}
\mathbf{f}=\sqrt{P}\frac{\mathbf{G}^H\mathrm{diag}\left(\mathbf{h}_\mathrm{r}\right)\mathbf{x}+\mathbf{h}}{\left\Vert\mathbf{G}^H\mathrm{diag}\left(\mathbf{h}_\mathrm{r}\right)\mathbf{x}+\mathbf{h}\right\Vert_2}.
\end{equation}

\emph{Remark 1:}  Note that matrix $\mathbf{R}$ in \eqref{eq5} is not PSD, and hence $\mathcal{P}_1$ is a non-convex problem. Furthermore, the element-wise UMCs $|v_i|=1$ are intrinsically non-convex, which is the main challenge in solving $\mathcal{P}_1$ optimally.
In summary, $\mathcal{P}_1$ is an NP-hard problem with the search dimension being $M+1$ \cite{lu2018argument}.

\emph{Remark 2:} The SDR approach was proposed to tackle $\mathcal{P}_1$ in \cite{8811733,tang2019joint}. In particular, an auxiliary optimization variable $\mathbf{V}=\mathbf{vv}^H$ was introduced to reformulate $\mathcal{P}_1$ as a semidefinite programming (SDP) problem with an additional rank-one constraint. By dropping the rank-one constraint and solving the SDP problem via standard convex optimization tools, the optimal solution for $\mathbf{V}$ can be obtained. However, there is no guarantee that the obtained solution $\mathbf{V}$ is a rank-one matrix. A Gaussian randomization approach was adopted, which ensures that the value of the objective function is asymptotically at least $\pi/4$ of the optimal value \cite{8811733}.
Therefore, the SDR approach can only provide an approximate solution for $\mathbf{v}$.

\emph{Remark 3:} The search space defined by the UMCs in $\mathcal{P}_1$ was identified as a complex circle manifold \cite{8855810}.
By translating the classical conjugate gradient descent methods in the Euclidean space to the  Riemannian manifold, a locally optimal solution was obtained for $\mathcal{P}_1$. To the best of the authors' knowledge, the manifold optimization-based algorithm achieves the highest spectral efficiency among all the existing approaches \cite{8855810}. In this paper, we  propose a BnB algorithm that yields the globally optimal solution of $\mathcal{P}_1$ and we study the degree of optimality of the manifold optimization.

\section{Branch-and-Bound Algorithm for IRS-Assisted MISO Wireless Communications}
The BnB algorithm was typically applied for solving NP-hard discrete and combinatorial optimization problems, and it has been recently adopted for solving continuous optimization problems \cite{8386661,lu2018argument}. 
The BnB algorithm is a systematic enumeration of candidate solutions by means of tree traversal. 
Each node in the search tree is associated with a set, which is a subset of the feasible set defined in the problem to be solved.
For each node, a subproblem is formulated with the corresponding subset, for which a lower bound and an upper bound are derived, in order to estimate the optimal solution of the subproblem.
In each iteration of the BnB algorithm, one node is selected according to the node selection rule, which is typically related to the bounds. Then, the selected node (associated set) is further branched into two child nodes (subsets). 
As the tree structure keeps growing, the feasible set is progressively partitioned into smaller subsets with improved objective values. 
In particular, following the BnB principles, we update the bounds of the selected subproblems in each iteration until convergence, i.e., the difference between the upper bound and lower bound goes to zero.
As suggested by the above discussion, there are three key factors in the BnB algorithm that have to be carefully designed, i.e., the chosen node for branching, the partition rule, and the bounding functions.

\subsection{Lower and Upper Bounds}
\begin{figure}[t]
	\centering\includegraphics[width=8cm]{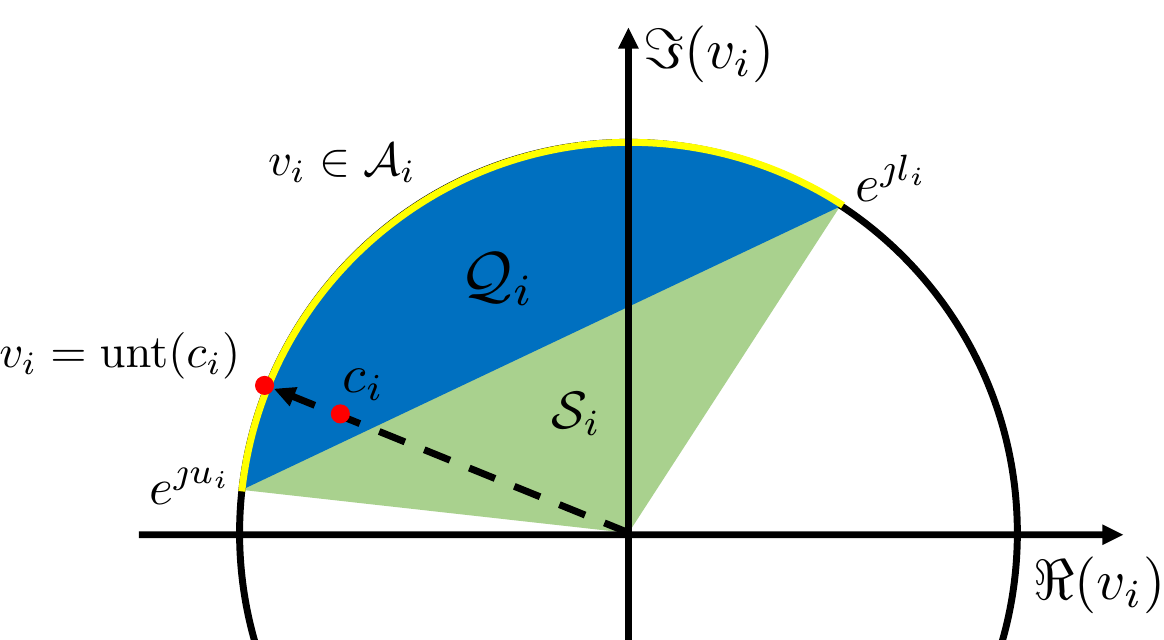}
	\caption{The feasible set $\mathcal{A}_i$ and the relaxed convex hull $\mathcal{Q}_i$  for each subproblem  $\mathcal{P}_2$.}
	\label{arc}
\end{figure}

The feasible set of $\mathcal{P}_1$ is the product of $M+1$ unit circles.
Therefore, the subset associated with any node in the search tree is denoted by $\mathcal{A}=\prod_{i=1}^{M+1}\mathcal{A}_i$, $i=1,2,\dots,M+1$. $\mathcal{A}_i$ denotes an arc whose endpoints are $e^{\jmath l_i}$ and $e^{\jmath u_i}$, cf. the yellow arc shown in Fig. \ref{arc}, where $l_i$ and $u_i$ are the limit points of the argument interval of the $i$-th element of $\mathbf{v}$.
Once one node is selected in each iteration of the BnB algorithm, the subproblem of $\mathcal{P}_1$ that needs to be solved is given by
\begin{equation}\mathcal{P}_2(\mathcal{A}):\quad
\begin{aligned}
&\underset{\mathbf{v}\in\mathbb{C}^{M+1}}{\mathrm{minimize}} && f(\mathbf{v})=\mathbf{v}^H\mathbf{Rv}\\
&\mathrm{subject\thinspace to}&&v_i\in\mathcal{A}_i,\quad \forall i.
\end{aligned}
\end{equation}
By defining $\mathbf{V}=\mathbf{vv}^H$, the BnB subproblem $\mathcal{P}_2$ is equivalent to
\begin{equation}\mathcal{P}_3\left(\mathcal{A}\right):\quad
\begin{aligned}
&\underset{\mathbf{v,V\succeq0}}{\mathrm{minimize}} &&\underline{g}(\mathbf{V})= \Tr\left(\mathbf{RV}\right)\\
&\mathrm{subject\thinspace to}&&v_i\in\mathcal{A}_i,\quad \forall i,\\
&&&\mathrm{Diag}\left(\mathbf{V}\right)=\mathbf{1}_{M+1},\\
&&&\mathbf{V}=\mathbf{vv}^H.
\end{aligned}
\end{equation}
For solving $\mathcal{P}_1$, according to the BnB principles, a lower bound and an upper bound need to be derived for subproblem $\mathcal{P}_3$.
More importantly, the tighter the bounds are, the faster the BnB algorithm converges \cite{8386661}. 
Therefore, the main task in this subsection is to find tight bounds for $\mathcal{P}_3$.

Note that the last equality constraint in $\mathcal{P}_3$ is basically a rank-one constraint, which is non-convex.
A commonly-adopted approach to derive a lower bound of $\mathcal{P}_3$ is to relax the non-convex constraint. 

\emph{Remark 4:} One may resort to the SDR approach in \cite{8811733,tang2019joint} to obtain a lower bound of $\mathcal{P}_3$. In this case, the rank-one constraint $\mathbf{V}=\mathbf{vv}^H$ is dropped. However, in this way, the first constraint of the optimization variable $\mathbf{v}$ is redundant, which causes the subproblems in all iterations of the BnB algorithm to be exactly the same. Therefore, the BnB algorithm does not converge, which indicates that the SDR approach is not applicable in the BnB framework.

In this paper, we first relax the rank-one constraint as $\mathbf{V}\succeq\mathbf{vv}^H$, which implies $|v_i|\le1$ for $i=1,\dots,M+1$. Combined with the first constraint in $\mathcal{P}_3$, the relaxed feasible set of $v_i$ is the circular sector determined by arc $\mathcal{A}_i$, cf. the region $\mathcal{Q}_i\cup\mathcal{S}_i$ in Fig. \ref{arc}. However, the circular sector is not the tightest convex relaxation of the feasible set $\mathcal{A}_i$ in $\mathcal{P}_3$. Instead, the circular segment enclosed by the arc $\mathcal{A}_i$ and the chord between $e^{\jmath l_i}$ and $e^{\jmath u_i}$ is the tightest convex relaxation, which is denoted as $\mathcal{Q}_i$ in Fig. \ref{arc}. Therefore, the following problem\footnote{As the relaxed solutions are no longer feasible solutions for $\mathcal{P}_3$, new denotations of the optimization variables, i.e., $\mathbf{c}$ and $\mathbf{C}$, are adopted here to distinguish the relaxed solutions from the original solutions $\mathbf{v}$ and $\mathbf{V}$.} 
needs to be solved in order to obtain a tight lower bound of $\mathcal{P}_3$:
\begin{equation}\mathcal{P}_4\left(\mathcal{Q}\right):\quad
\begin{aligned}
&\underset{\mathbf{c,C\succeq0}}{\mathrm{minimize}} && \underline{g}(\mathbf{C})=\Tr\left(\mathbf{RC}\right)\\
&\mathrm{subject\thinspace to}&&c_i\in\mathcal{Q}_i,\quad \forall i,\\
&&&\mathrm{Diag}\left(\mathbf{C}\right)=\mathbf{1}_{M+1},\\
&&&\mathbf{C}\succeq\mathbf{cc}^H,
\end{aligned}
\end{equation}
where  $\mathcal{Q}=\prod_{i=1}^{M+1}\mathcal{Q}_i$.
The first constraint in $\mathcal{P}_4$ can be recast as 
\begin{equation}\label{con1}
\Re\left(\mathbf{a}^*\circ\mathbf{c}\right)\ge\cos\left(\frac{\mathbf{u-l}}{2}\right),
\end{equation}
where $\mathbf{l}=\left[l_1,\dots,l_{M+1}\right]^T$, $\mathbf{u}=\left[u_1,\dots,u_{M+1}\right]^T$, and $a_i=e^{\jmath\frac{{u_i+l_i}}{2}}$ for $i=1,2,\dots,M+1$. In addition, according to the Schur complement, the last constraint in $\mathcal{P}_4$ is equivalent to the following linear matrix inequality
\begin{equation}\label{con2}
\begin{bmatrix}
1&\mathbf{c}^H\\
\mathbf{c}&\mathbf{C}
\end{bmatrix}\succeq0.
\end{equation}
Note that both \eqref{con1} and \eqref{con2} are convex constraints. Therefore, $\mathcal{P}_4$ can be efficiently solved by standard convex program solvers such as CVX \cite{grant2009cvx}. 

On the other hand, an upper bound of $\mathcal{P}_3$ can be obtained by a feasible solution for $\mathcal{P}_3$. In this paper, we project the optimal solution of $\mathcal{P}_4$, i.e., $\mathbf{c}$, onto the feasible set of $\mathcal{P}_3$, namely, $\mathcal{A}$. In particular,  the upper bound is given by $f({{\mathbf{v}}})$, where ${\mathbf{v}}$ is a feasible solution given by
\begin{equation}\label{round}
{\mathbf{v}}=\mathrm{unt}\left(\mathbf{c}\right).
\end{equation}

\begin{algorithm}[t]
	\caption{BnB Algorithm for Solving $\mathcal{P}_1$}
	\begin{algorithmic}[1]
		\STATE Initialize $\mathcal{A}^0$ as the product of $M+1$ unit circles. Solve $\mathcal{P}_4\left(\mathcal{Q}^0\right)$ for its optimal solution $\left\{\mathbf{c}^0,\mathbf{C}^0\right\}$, and compute the feasible solution ${\mathbf{v}}^0$ according to \eqref{round}. 
		Add the node associated with $\left\{\mathcal{A}^0,\mathbf{c}^0,\mathbf{C}^0\right\}$ to the search tree $\mathcal{T}$.
		Set convergence tolerance $\epsilon$ and iteration index $t=0$, 
		\REPEAT 
		\STATE $t\leftarrow t+1$;
		\STATE Select the node associated with $\left\{\mathcal{A}^t,\mathbf{c}^t,\mathbf{C}^t\right\}$ such that $\underline{g}(\mathbf{C}^t)$ is the smallest lower bound among all the nodes;
		\STATE Partition the feasible set associated with the selected node into two subsets, $\mathcal{A}^t_\mathrm{l}$ and $\mathcal{A}^t_\mathrm{r}$, according to \eqref{eq20};
		\STATE Solve $\mathcal{P}_4\left(\mathcal{Q}^t_\mathrm{l}\right)$ for its optimal solution $\left\{\mathbf{c}^t_\mathrm{l},\mathbf{C}^t_\mathrm{l}\right\}$, and compute the feasible solution ${\mathbf{v}}^t_\mathrm{l}$ according to \eqref{round};
		\STATE Solve $\mathcal{P}_4\left(\mathcal{Q}^t_\mathrm{r}\right)$ for its optimal solution $\left\{\mathbf{c}^t_\mathrm{r},\mathbf{C}^t_\mathrm{r}\right\}$, and compute the feasible solution ${\mathbf{v}}^t_\mathrm{r}$ according to \eqref{round};
		\STATE Add the two partitioned nodes associated with $\left\{\mathcal{A}^t_\mathrm{l},\mathbf{c}^t_\mathrm{l},\mathbf{C}^t_\mathrm{l}\right\}$ and $\left\{\mathcal{A}^t_\mathrm{r},\mathbf{c}^t_\mathrm{r},\mathbf{C}^t_\mathrm{r}\right\}$ to $\mathcal{T}$;
		\STATE Update $U^t$ and $L^t$ as the smallest upper bound $f({\mathbf{v}}^t)$
		and lower bound $\underline{g}\left(\mathbf{C}^t\right)$ in $\mathcal{T}$, respectively;
		
		\UNTIL $\frac{U^t-L^t}{L^t}\le\epsilon$
		\STATE Update the optimal solution of $\mathcal{P}_1$ as $\mathbf{v}^\star={\mathbf{v}}^t$.
	\end{algorithmic}
\end{algorithm}

\subsection{Node Selection and Partition Rules}
In each iteration of the BnB algorithm, a node in the search tree is selected to be further branched. In this paper, we select the node associated with the smallest lower bound, and partition its corresponding feasible set $\mathcal{A}$ according to the Euclidean distance between the solution of $\mathcal{P}_4$, i.e., $\mathbf{c}$, and its projected solution ${\mathbf{v}}$. In particular, we equally partition $\mathcal{A}_{i^\star}$ and keep $\mathcal{A}_i$ for $ i\ne i^\star$
unchanged, where
\begin{equation}\label{eq20}
i^\star=\arg\underset{i}{\max}\quad|{c_i}-{v}_i|.
\end{equation}
According to \cite[Lemma 4]{lu2018argument},  with the three key elements presented in this section, i.e., obtained bounds, the node selection rule, and the node partition rule, the BnB algorithm is guaranteed to converge to an $\epsilon$-optimal solution, where $\epsilon$ is the convergence tolerance.
As is well known in the literature, the worst-case computational complexity grows exponentially with $M+1$, where $M$ is the number of IRS elements.
The proposed BnB algorithm is summarized in \textbf{Algorithm 1}.

\section{Simulation Results}
In this section, we evaluate the performance of the proposed BnB algorithm. The carrier center frequency is $2.4$ GHz. All channels are assumed to be independent Rayleigh fading, and the path loss exponent is set to $3$ with reference distance $10$ m. 
The AP-user distance, IRS-user distance, and AP-IRS distance are denoted by $r_\mathrm{Au}$, $r_\mathrm{Iu}$, and $r_\mathrm{AI}$, respectively.
The total transmit power is $P=10$ dBm while the noise power at the user is set to $\sigma^2=-90$ dBm. All simulation results in this section are averaged over 1000 channel realizations. 

\subsection{Convergence of the Proposed BnB Algorithm}
\begin{figure}[t]
	\centering\includegraphics[width=8.5cm]{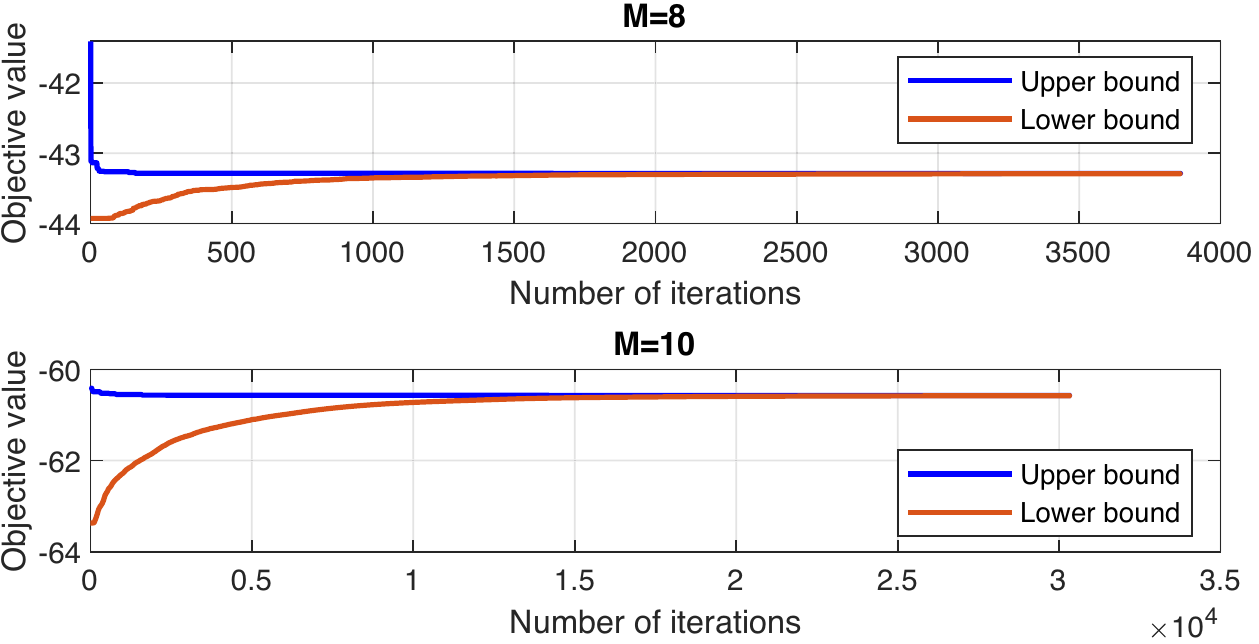}
	\caption{Convergence of the proposed BnB algorithm for different values of $M$. The system parameters are set as $\Nt=4$, $r_\mathrm{Au}=50$ m, $r_\mathrm{AI}=30$ m,  $r_\mathrm{Iu}=20$ m, and $\epsilon=10^{-5}$.}
	\label{fig1}
\end{figure}
In Fig. \ref{fig1}, we investigate the convergence of the proposed BnB algorithm for different numbers of IRS elements, $M$. 
The convergence tolerance of  \textbf{Algorithm 1} is set to $\epsilon=10^{-5}$.
As can be observed from Fig. \ref{fig1}, the upper bound $U^t$ and lower bound $L^t$ in the proposed BnB algorithm monotonically converge to the same value for both considered values of $M$. 
In particular, the number of iterations needed for achieving the convergence tolerance is around $3800$ for  $M=8$.
In contrast, for the case with $M=10$ IRS elements, the proposed BnB algorithm needs significantly more iterations for convergence, i.e., over $3\times10^4$ iterations.
Note that adding only two IRS elements results in a tremendous increase in the number of iterations. This is because the size of the search tree grows exponentially with $M+1$. Furthermore, after convergence, the normalized objective value for $M=10$ is lower than that for $M=8$. This indicates that deploying more IRS elements is beneficial for IRS-assisted MISO communication systems.

\subsection{Massive MIMO or Large-Scale IRS?}
\begin{figure}[t]
	\centering\includegraphics[width=8.5cm]{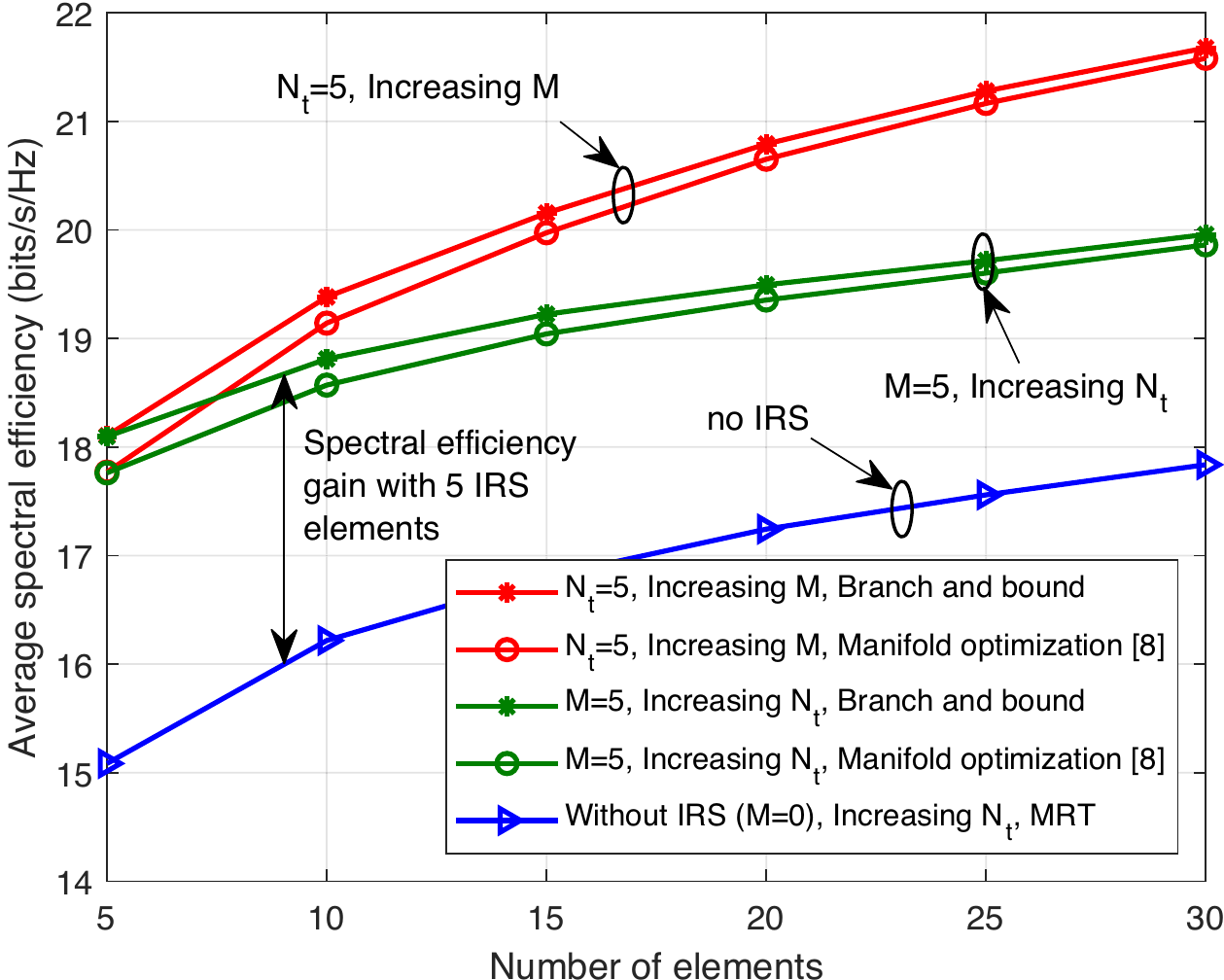}
	\caption{Average spectral efficiency achieved for different values of $M$ and $\Nt$ when $r_\mathrm{AI} =r_\mathrm{Au} =50$ m and $r_\mathrm{Iu} =20$ m.}
	\label{fig2}
\end{figure}
For conventional wireless communications systems, deploying large-scale antenna arrays at the transceivers is an effective way to boost the network capacity. 
The blue curve in Fig. \ref{fig2} illustrates this effect assuming that there is no IRS deployed in the network, and optimal MRT beamforming is adopted to align the beam to the direct channel $\mathbf{h}$, i.e., $\mathbf{f}=\sqrt{P}\mathbf{h}/\left\Vert\mathbf{h}\right\Vert_2$. For the IRS-assisted system considered in this paper, the red curves in Fig. \ref{fig2} represent the spectral efficiency achieved for increasing values of $M$, while keeping the transmit antenna array size as $\Nt = 5$.
On the other hand, the green curves depict the spectral efficiency achieved for increasing numbers of transmit antenna elements, $\Nt$, when using a $5$-element IRS.
We observe that both considered IRS-assisted systems significantly outperform the MRT strategy without IRSs, which confirms the effectiveness of incorporating IRSs into wireless communication systems.

Fig. \ref{fig2} clearly shows that increasing the number of IRS elements is more efficient than enlarging the antenna size at the AP in terms of improving spectral efficiency. 
Furthermore, additional RF chains and power amplifiers need to be deployed for driving the increasing number of antenna elements, which leads to a more energy-consuming wireless system compared to the deployment of large-scale passive IRSs. Therefore, we conclude that IRS-assisted wireless systems are more spectral- and energy-efficient than conventional wireless systems.

As the proposed BnB algorithm is guaranteed to converge to the optimal solution of $\mathcal{P}_1$, it can be regarded as a performance benchmark for existing suboptimal algorithms. 
As discussed in Remark 3, the manifold optimization-based algorithm proposed in \cite{8855810} achieves the highest spectral efficiency among the existing approaches for beamforming in IRS-assisted MISO communication systems.
As can be observed in Fig. \ref{fig2}, by taking the proposed BnB algorithm as the benchmark, the manifold optimization-based algorithm achieves a near-optimal solution, especially in large-scale wireless systems.
Therefore, with the help of the proposed BnB algorithm, the manifold optimization in \cite{8855810} is shown to be an efficient algorithm for designing large-scale IRS-assisted MISO communication systems.

\section{Conclusions}
In this paper, we investigated the joint design of the active beamformer at the AP and the passive beamformer at the IRS in an IRS-assisted single-user MISO wireless communication system.
A BnB algorithm was proposed for tackling the UMCs, which are the main obstacles for optimizing the beamformers. It is the first globally optimal algorithm developed for IRS-assisted MISO systems in the literature. 
Simulation results revealed the substantial potential of IRSs for establishing high-speed green communication networks. Moreover, by taking the proposed BnB algorithm as the performance benchmark, low-complexity manifold optimization was shown to be a near-optimal algorithm for large-scale IRS-aided wireless systems.
%


\bibliographystyle{IEEEtran}
%
\bibliography{bare_conf}

\end{document}